\documentclass[11pt]{article}
\usepackage{graphicx}
\usepackage{dcolumn}     % Align table columns on decimal point
\usepackage{bm}

\newcommand{\BABARPubYear}    {06}
\newcommand{\BABARConfNumber} {211}
\newcommand{\SLACPubNumber} {12244}

\input babarsym
\usepackage{relsize}
\def\babar{\mbox{\slshape B\kern-0.1em{\smaller A}\kern-0.1em
    B\kern-0.1em{\smaller A\kern-0.2em R}}}

\def\en         {\ensuremath{e^-}\xspace}   % electron negative (\em is taken)
\def\ep         {\ensuremath{e^+}\xspace}
 
\def\epem       {\ensuremath{e^+e^-}\xspace}

\def\mup        {\ensuremath{\mu^+}\xspace}
\def\mun        {\ensuremath{\mu^-}\xspace} % muon negative (\mum is taken)
\def\mumu       {\ensuremath{\mu^+\mu^-}\xspace}

\def\taup       {\ensuremath{\tau^+}\xspace}
\def\taum       {\ensuremath{\tau^-}\xspace}

\def\ellm       {\ensuremath{\ell^-}\xspace}

\def\ellell     {\ensuremath{\ell^+ \ell^-}\xspace}

\def\nub        {\ensuremath{\overline{\nu}}\xspace}
\def\nunub      {\ensuremath{\nu{\overline{\nu}}}\xspace}
\def\nub        {\ensuremath{\overline{\nu}}\xspace}
\def\nunub      {\ensuremath{\nu{\overline{\nu}}}\xspace}
\def\nue        {\ensuremath{\nu_e}\xspace}
\def\nueb       {\ensuremath{\nub_e}\xspace}

\def\num        {\ensuremath{\nu_\mu}\xspace}
\def\numb       {\ensuremath{\nub_\mu}\xspace}

\def\nut        {\ensuremath{\nu_\tau}\xspace}
\def\nutb       {\ensuremath{\nub_\tau}\xspace}

\def\nul        {\ensuremath{\nu_\ell}\xspace}
\def\nulb       {\ensuremath{\nub_\ell}\xspace}

\def\g     {\ensuremath{\gamma}\xspace}
\def\gaga  {\ensuremath{\gamma\gamma}\xspace}  %% changed from \gg, which is >>
\def\ggstar{\ensuremath{\gamma\gamma^*}\xspace}

\def\piz   {\ensuremath{\pi^0}\xspace}

\def\ppz   {\ensuremath{\pi^0\pi^0}\xspace}
\def\pip   {\ensuremath{\pi^+}\xspace}
\def\pim   {\ensuremath{\pi^-}\xspace}
\def\pipi  {\ensuremath{\pi^+\pi^-}\xspace}

\def\pimp  {\ensuremath{\pi^\mp}\xspace}

\def\kaon  {\ensuremath{K}\xspace}
\def\Kbar  {\kern 0.2em\overline{\kern -0.2em K}{}\xspace}

\def\Kz    {\ensuremath{K^0}\xspace}
\def\Kzb   {\ensuremath{\Kbar^0}\xspace}
\def\KzKzb {\ensuremath{\Kz \kern -0.16em \Kzb}\xspace}
\def\Kp    {\ensuremath{K^+}\xspace}
\def\Km    {\ensuremath{K^-}\xspace}
\def\Kpm   {\ensuremath{K^\pm}\xspace}

\def\KpKm  {\ensuremath{\Kp \kern -0.16em \Km}\xspace}
\def\KS    {\ensuremath{K^0_{\scriptscriptstyle S}}\xspace} 
\def\KL    {\ensuremath{K^0_{\scriptscriptstyle L}}\xspace} 
\def\Kstarz  {\ensuremath{K^{*0}}\xspace}

\def\Kstar   {\ensuremath{K^*}\xspace}

\def\Kstarp  {\ensuremath{K^{*+}}\xspace}

\def\Dbar    {\kern 0.2em\overline{\kern -0.2em D}{}\xspace}
\def\Db      {\ensuremath{\Dbar}\xspace}
\def\Dz      {\ensuremath{D^0}\xspace}
\def\Dzb     {\ensuremath{\Dbar^0}\xspace}
\def\DzDzb   {\ensuremath{\Dz {\kern -0.16em \Dzb}}\xspace}
\def\Dp      {\ensuremath{D^+}\xspace}
\def\Dm      {\ensuremath{D^-}\xspace}

\def\DpDm    {\ensuremath{\Dp {\kern -0.16em \Dm}}\xspace}
\def\Dstar   {\ensuremath{D^*}\xspace}

\def\Dstarz  {\ensuremath{D^{*0}}\xspace}

\def\Dstarp  {\ensuremath{D^{*+}}\xspace}
\def\Dstarm  {\ensuremath{D^{*-}}\xspace}

\def\B       {\ensuremath{B}\xspace}
\def\Bbar    {\kern 0.18em\overline{\kern -0.18em B}{}\xspace}
\def\Bb      {\ensuremath{\Bbar}\xspace}
\def\BB      {\ensuremath{\B {\kern -0.16em \Bb}}\xspace}
\def\Bz      {\ensuremath{B^0}\xspace}
\def\Bzb     {\ensuremath{\Bbar^0}\xspace}
\def\BzBzb   {\ensuremath{\Bz {\kern -0.16em \Bzb}}\xspace}

\def\Bu      {\ensuremath{B^+}\xspace}
\def\Bub     {\ensuremath{B^-}\xspace}

\def\Bpm     {\ensuremath{B^\pm}\xspace}

\def\BpBm    {\ensuremath{\Bu {\kern -0.16em \Bub}}\xspace}

\def\BorBbar    {\kern 0.18em\optbar{\kern -0.18em B}{}\xspace}
\def\DorDbar    {\kern 0.18em\optbar{\kern -0.18em D}{}\xspace}
\def\KorKbar    {\kern 0.18em\optbar{\kern -0.18em K}{}\xspace}

\def\jpsi     {\ensuremath{{J\mskip -3mu/\mskip -2mu\psi\mskip 2mu}}\xspace}

\mathchardef\Upsilon="7107
\def\Y#1S{\ensuremath{\Upsilon{(#1S)}}\xspace}% no space before {...}!

\def\FourS {\Y4S}

\mathchardef\Deltares="7101
\mathchardef\Xi="7104
\mathchardef\Lambda="7103
\mathchardef\Sigma="7106
\mathchardef\Omega="710A

\def\Deltabar{\kern 0.25em\overline{\kern -0.25em \Deltares}{}\xspace}
\def\Lbar{\kern 0.2em\overline{\kern -0.2em\Lambda\kern 0.05em}\kern-0.05em{}\xspace}
\def\Sigbar{\kern 0.2em\overline{\kern -0.2em \Sigma}{}\xspace}
\def\Xibar{\kern 0.2em\overline{\kern -0.2em \Xi}{}\xspace}
\def\Obar{\kern 0.2em\overline{\kern -0.2em \Omega}{}\xspace}
\def\Nbar{\kern 0.2em\overline{\kern -0.2em N}{}\xspace}
\def\Xb{\kern 0.2em\overline{\kern -0.2em X}{}\xspace}

\def\X {\ensuremath{X}\xspace}

\def\BR         {{\ensuremath{\cal B}\xspace}}

\def\bpsikl     {\ensuremath{\Bz \to \jpsi \KL}\xspace}

\def\Bzbtox     {\ensuremath{\Bzb \to \X}\xspace}

\def\invfb   {\ensuremath{\mbox{\,fb}^{-1}}\xspace}

\def\mus  {\ensuremath{\rm \,\mus}\xspace}

\def\mus        {\ensuremath{\,\mu{\rm s}}\xspace}    %% microsecond
\def\degrees{\ensuremath{^{\circ}}\xspace}
\def\to                 {\ensuremath{\rightarrow}\xspace}

\def\pep2{PEP-II}
\def\BF{$B$ Factory}

\def\gsim{{~\raise.15em\hbox{$>$}\kern-.85em
          \lower.35em\hbox{$\sim$}~}\xspace}
\def\lsim{{~\raise.15em\hbox{$<$}\kern-.85em
          \lower.35em\hbox{$\sim$}~}\xspace}

\def\epstag  {\ensuremath{\varepsilon_{\rm tag}}\xspace}

\def\Vcb  {\ensuremath{|V_{cb}|}\xspace}

\def\jetset74   {\mbox{\tt Jetset \hspace{-0.5em}7.\hspace{-0.2em}4}\xspace}
\usepackage{relsize}

\def\Mnu{\ensuremath{{\cal M}^2}}

\long\def\inst#1{\par\nobreak\kern 4pt\nobreak
    {\it #1}\par\vskip 10pt plus 3pt minus 3pt}

\RequirePackage{xspace}

\usepackage{relsize}
\def\babar{\mbox{\slshape B\kern-0.1em{\smaller A}\kern-0.1em
    B\kern-0.1em{\smaller A\kern-0.2em R}}}

\def\en         {\ensuremath{e^-}\xspace}   % electron negative (\em is taken)
\def\ep         {\ensuremath{e^+}\xspace}
 
\def\epem       {\ensuremath{e^+e^-}\xspace}

\def\mup        {\ensuremath{\mu^+}\xspace}
\def\mun        {\ensuremath{\mu^-}\xspace} % muon negative (\mum is taken)
\def\mumu       {\ensuremath{\mu^+\mu^-}\xspace}

\def\taup       {\ensuremath{\tau^+}\xspace}
\def\taum       {\ensuremath{\tau^-}\xspace}

\def\ellm       {\ensuremath{\ell^-}\xspace}

\def\ellell     {\ensuremath{\ell^+ \ell^-}\xspace}

\def\nub        {\ensuremath{\overline{\nu}}\xspace}
\def\nunub      {\ensuremath{\nu{\overline{\nu}}}\xspace}
\def\nub        {\ensuremath{\overline{\nu}}\xspace}
\def\nunub      {\ensuremath{\nu{\overline{\nu}}}\xspace}
\def\nue        {\ensuremath{\nu_e}\xspace}
\def\nueb       {\ensuremath{\nub_e}\xspace}

\def\num        {\ensuremath{\nu_\mu}\xspace}
\def\numb       {\ensuremath{\nub_\mu}\xspace}

\def\nut        {\ensuremath{\nu_\tau}\xspace}
\def\nutb       {\ensuremath{\nub_\tau}\xspace}

\def\nul        {\ensuremath{\nu_\ell}\xspace}
\def\nulb       {\ensuremath{\nub_\ell}\xspace}

\def\g     {\ensuremath{\gamma}\xspace}
\def\gaga  {\ensuremath{\gamma\gamma}\xspace}  %% changed from \gg, which is >>
\def\ggstar{\ensuremath{\gamma\gamma^*}\xspace}

\def\piz   {\ensuremath{\pi^0}\xspace}

\def\ppz   {\ensuremath{\pi^0\pi^0}\xspace}
\def\pip   {\ensuremath{\pi^+}\xspace}
\def\pim   {\ensuremath{\pi^-}\xspace}
\def\pipi  {\ensuremath{\pi^+\pi^-}\xspace}

\def\pimp  {\ensuremath{\pi^\mp}\xspace}

\def\kaon  {\ensuremath{K}\xspace}
\def\Kbar  {\kern 0.2em\overline{\kern -0.2em K}{}\xspace}

\def\Kz    {\ensuremath{K^0}\xspace}
\def\Kzb   {\ensuremath{\Kbar^0}\xspace}
\def\KzKzb {\ensuremath{\Kz \kern -0.16em \Kzb}\xspace}
\def\Kp    {\ensuremath{K^+}\xspace}
\def\Km    {\ensuremath{K^-}\xspace}
\def\Kpm   {\ensuremath{K^\pm}\xspace}

\def\KpKm  {\ensuremath{\Kp \kern -0.16em \Km}\xspace}
\def\KS    {\ensuremath{K^0_{\scriptscriptstyle S}}\xspace} 
\def\KL    {\ensuremath{K^0_{\scriptscriptstyle L}}\xspace} 
\def\Kstarz  {\ensuremath{K^{*0}}\xspace}

\def\Kstar   {\ensuremath{K^*}\xspace}

\def\Kstarp  {\ensuremath{K^{*+}}\xspace}

\def\Dbar    {\kern 0.2em\overline{\kern -0.2em D}{}\xspace}
\def\Db      {\ensuremath{\Dbar}\xspace}
\def\Dz      {\ensuremath{D^0}\xspace}
\def\Dzb     {\ensuremath{\Dbar^0}\xspace}
\def\DzDzb   {\ensuremath{\Dz {\kern -0.16em \Dzb}}\xspace}
\def\Dp      {\ensuremath{D^+}\xspace}
\def\Dm      {\ensuremath{D^-}\xspace}

\def\DpDm    {\ensuremath{\Dp {\kern -0.16em \Dm}}\xspace}
\def\Dstar   {\ensuremath{D^*}\xspace}

\def\Dstarz  {\ensuremath{D^{*0}}\xspace}

\def\Dstarp  {\ensuremath{D^{*+}}\xspace}
\def\Dstarm  {\ensuremath{D^{*-}}\xspace}

\def\B       {\ensuremath{B}\xspace}
\def\Bbar    {\kern 0.18em\overline{\kern -0.18em B}{}\xspace}
\def\Bb      {\ensuremath{\Bbar}\xspace}
\def\BB      {\ensuremath{\B {\kern -0.16em \Bb}}\xspace}
\def\Bz      {\ensuremath{B^0}\xspace}
\def\Bzb     {\ensuremath{\Bbar^0}\xspace}
\def\BzBzb   {\ensuremath{\Bz {\kern -0.16em \Bzb}}\xspace}

\def\Bu      {\ensuremath{B^+}\xspace}
\def\Bub     {\ensuremath{B^-}\xspace}

\def\Bpm     {\ensuremath{B^\pm}\xspace}

\def\BpBm    {\ensuremath{\Bu {\kern -0.16em \Bub}}\xspace}

\def\BorBbar    {\kern 0.18em\optbar{\kern -0.18em B}{}\xspace}
\def\DorDbar    {\kern 0.18em\optbar{\kern -0.18em D}{}\xspace}
\def\KorKbar    {\kern 0.18em\optbar{\kern -0.18em K}{}\xspace}

\def\jpsi     {\ensuremath{{J\mskip -3mu/\mskip -2mu\psi\mskip 2mu}}\xspace}

\mathchardef\Upsilon="7107
\def\Y#1S{\ensuremath{\Upsilon{(#1S)}}\xspace}% no space before {...}!

\def\FourS {\Y4S}

\mathchardef\Deltares="7101
\mathchardef\Xi="7104
\mathchardef\Lambda="7103
\mathchardef\Sigma="7106
\mathchardef\Omega="710A

\def\Deltabar{\kern 0.25em\overline{\kern -0.25em \Deltares}{}\xspace}
\def\Lbar{\kern 0.2em\overline{\kern -0.2em\Lambda\kern 0.05em}\kern-0.05em{}\xspace}
\def\Sigbar{\kern 0.2em\overline{\kern -0.2em \Sigma}{}\xspace}
\def\Xibar{\kern 0.2em\overline{\kern -0.2em \Xi}{}\xspace}
\def\Obar{\kern 0.2em\overline{\kern -0.2em \Omega}{}\xspace}
\def\Nbar{\kern 0.2em\overline{\kern -0.2em N}{}\xspace}
\def\Xb{\kern 0.2em\overline{\kern -0.2em X}{}\xspace}

\def\X {\ensuremath{X}\xspace}

\def\BR         {{\ensuremath{\cal B}\xspace}}

\def\bpsikl     {\ensuremath{\Bz \to \jpsi \KL}\xspace}

\def\Bzbtox     {\ensuremath{\Bzb \to \X}\xspace}

\def\invfb   {\ensuremath{\mbox{\,fb}^{-1}}\xspace}

\def\mus  {\ensuremath{\rm \,\mus}\xspace}

\def\mus        {\ensuremath{\,\mu{\rm s}}\xspace}    %% microsecond
\def\degrees{\ensuremath{^{\circ}}\xspace}
\def\to                 {\ensuremath{\rightarrow}\xspace}
\def\pep2{PEP-II}
\def\BF{$B$ Factory}

\def\gsim{{~\raise.15em\hbox{$>$}\kern-.85em
          \lower.35em\hbox{$\sim$}~}\xspace}
\def\lsim{{~\raise.15em\hbox{$<$}\kern-.85em
          \lower.35em\hbox{$\sim$}~}\xspace}

\def\epstag  {\ensuremath{\varepsilon_{\rm tag}}\xspace}

\def\Vcb  {\ensuremath{|V_{cb}|}\xspace}

\def\jetset74   {\mbox{\tt Jetset \hspace{-0.5em}7.\hspace{-0.2em}4}\xspace}
\usepackage{relsize}

\def\Mnu{\ensuremath{{\cal M}^2}}

\begin{document}
{\pagestyle{empty}

\begin{flushright}
%BAD827 Version 14\\
\babar-CONF-\BABARPubYear/\BABARConfNumber \\
%\babar-PUB-\BABARPubYear/\BABARPubNumber \\
SLAC-PUB-\SLACPubNumber \\
%hep-ex/\LANLNumber \\
December 2006 \\
\end{flushright}

\par\vskip 5cm

% Title of the paper
\begin{center}
\Large \bf \boldmath 
Measurement of the Absolute Branching Fraction of $\Dz \rightarrow \Km \pip$
\end{center}
\bigskip

\begin{center}
\large Romulus Godang\footnote{E-mail: godang@slac.stanford.edu} (on behalf of the \babar\ Collaboration)\\
\vspace{0.15cm}
{\em Department of Physics and Astronomy\\ 
University of Mississippi-Oxford, University, MS 38677}\\
\vspace{0.15cm}
and\\
\vspace{0.15cm}
{\em Stanford Linear Accelerator Center, Stanford University, Stanford, CA 94309.}
\mbox{ }\\
\end{center}
\bigskip \bigskip

% Abstract
\begin{center}
\large
\end{center}
We measure the absolute branching fraction for $\Dz \rightarrow \Km \pip$ 
using partial reconstruction of $\Bzb \rightarrow D^{*+} X \ell^{-} \bar{\nu}_{\ell}$ decays. 
Only the charged lepton and the soft pion from the decay $D^{*+} \rightarrow D^{0} \pi^{+}$ are used.
Based on a data sample of 230 million \BB pairs collected at the $\Upsilon(4S)$
resonance with the \babar\ detector at the PEP-II asymmetric-energy $B$
Factory at SLAC, we obtain  ${\cal B}(\Dz \rightarrow \Km \pip) = 
(4.025 \pm 0.038 \pm 0.098)\%$, where the first error is statistical and the second 
error is systematic. 
\vspace{0.7cm}

\begin{center}
Presented at Joint Meeting of Pacific Region Particle Physics Communities\\
DPF 2006 + JPS 2006, Honolulu, Hawaii, October 29 - November 03, 2006.
\end{center}

\vspace{1.0cm}
\begin{center}
{\em Stanford Linear Accelerator Center, Stanford University, 
Stanford, CA 94309} \\ \vspace{0.1cm}\hrule\vspace{0.1cm}
Work supported in part by the U.S. Department of Energy contracts\\
DE-AC02-76SF00515 and DE-FG05-91ER40622.
\end{center}

\newpage

} % end of pagestyle{empty}

% The body of the paper starts here

\section{Introduction}

The decay $\Dz \rightarrow \Km \pip$ is used as a  
normalizing mode in many measurements of $D$ and $B$-mesons decay branching fractions.
A precise measurement of the value of  ${\cal B}(\Dz \rightarrow \Km \pip)$
improves our knowledge of $D$ and $B$-meson properties, and of fundamental parameters
of the Standard Model, such as the Cabibbo-Kobayashi-Maskawa matrix element \Vcb~\cite{CKM},\cite{panic05}. 
The CLEO-c Collaboration has recently published a result achieving a few percent 
accuracy~\cite{cleoc05}.
We present here a measurement of comparable precision
based on partial reconstruction of the decay $\Bzb \rightarrow D^{*+} X \ell^{-} \bar{\nu}_{\ell}$,
inspired by a similar measurement performed by CLEO~\cite{cleo98}.
This partial reconstruction method was introduced by ARGUS~\cite{argus94} 
to measure \BzBzb\ mixing, and has been exploited by 
DELPHI~\cite{delphi97}, OPAL~\cite{opal00}, and CLEO~\cite{godang02} to measure several
$B$-meson properties. \babar\ applied this technique to measure the \Bzb\-meson lifetime~\cite{franco02}, 
the branching fraction of $\Upsilon(4S) \rightarrow \BzBzb$~\cite{godang05}, and for 
an improved determination of the \Bzb lifetime and \BzBzb oscillation frequency~\cite{franco06}.

\section{Dataset and Selection}

The data sample used in this analysis consists of 210\invfb, corresponding to 230 million \BB pairs, 
collected at the $\Y4S$ resonance (on-resonance) and 22\invfb collected 40\mev 
below the resonance (off-resonance) by the \babar\ detector. The off-resonance events are 
used to subtract the non-\BB\ background (continuum) from light quark and lepton 
production. A simulated sample of $\BB$ events with integrated luminosity equivalent to approximately 
five times the data is used for efficiency computation and background studies.

A detailed description of the \babar\ detector and the algorithms used
for particles reconstruction and identification is provided
elsewhere~\cite{babar_nim}. 
We summarize here only the performances more 
relevant to this measurement. 
High-momentum particles are reconstructed by matching hits in 
the silicon vertex tracker (SVT) with track elements in the drift chamber (DCH). 
Lower momentum tracks, which do not leave signals on many wires in the DCH due 
to the bending induced by a magnetic field, are reconstructed in the SVT.
Charged hadron identification is performed combining the  measurements of 
the energy deposition in the SVT and in the DCH with the information from the
Cherenkov detector (DIRC).
Electrons are identified by the ratio of the track momentum to the
associated energy deposited in the calorimeter (EMC), the transverse
profile of the shower, the energy loss in the DCH, and the Cherenkov angle in the DIRC.
Muons are identified in the instrumented flux return (IFR), composed
of resistive plate chambers and layers of iron.
Muon candidates are required to have a path length and hit
distribution in the IFR and energy deposition in the EMC consistent 
with that expected for a minimum-ionizing particle.

\section{Analysis Technique}

We preselect a sample of hadronic events with at least four charged tracks.
To reduce continuum background, we require that the ratio of the 2$^{nd}$ to the
0$^{th}$ order Fox-Wolfram~\cite{wolfram} variable be $R_2<0.6$.
We then select the sample of partially reconstructed $B$ mesons in the channel 
$\Bzb \rightarrow D^{*+} X \ell^{-} \bar{\nu}_{\ell}$, 
searching for the charged lepton ($\ell = e,\,\mu$) and the low momentum 
pion (soft pion), $\pi^+_{s}$, daughter from the decay $D^{*+}\to \Dz \pi^+_{s}$
\footnote{The inclusion of charge-conjugate reactions is implied throughout this paper.}.
This sample of events is referred to as ``inclusive sample''.
Using the conservation of momentum and energy, in the presence of an undetected neutrino, 
the neutrino invariant mass squared is calculated as
\begin{eqnarray}
\Mnu \equiv (E_{\mbox{\rm beam}}-E_{{D^*}} - 
E_{\ell})^2-({\bf{p}}_{{D^*}} + {\bf{p}}_{\ell})^2\,,
\label{eqn:mms}
\end{eqnarray}
where $E_{\mbox{\rm beam}}$ is half the center-of-mass energy and $E_{\ell}~(E_{{D^*}})$ 
and ${\bf{p}}_{\ell}~({\bf{p}}_{{D^*}})$ are respectively the center-of-mass energy and momentum 
of the lepton (the $D^*$ meson).  Since the $B$ momentum (${\bf{p}}_{B}$) is sufficiently small compared to 
$|{\bf{p}}_{\ell}|$ and $|{\bf{p}}_{D^*}|$, we set ${\bf{p}}_{B}$ = 0.
As a consequence of the limited phase space available in the $D^{*+}$
decay, the soft pion is emitted nearly at rest in the $D^{*+}$ rest frame.
The $D^{*+}$ four-momentum can therefore be computed by approximating 
its direction as that of the soft pion, and parameterizing its momentum as 
a linear function of the soft-pion momentum.
The lepton momentum must be in the range $1.4 < p_{\ellm} < 2.3 \gevc$ and 
the soft pion candidate must satisfy $60 < p_{\pi^{+}_{s}} < 190 \mevc$ in the $\epem$ center-of-mass 
frame. The two tracks must be consistent with originating from a common vertex, constrained to the 
beam-spot in the plane transverse to the beam axis. Finally, we combine $p_{\ellm}$, $p_{\pi^{+}_{s}}$ and the
probability from the vertex fit in a likelihood ratio variable ($\chi$), optimized to reject \BB\
background. If we find more than one candidate in the event, we choose the one with
the largest value of $\chi$.
We select pairs of tracks with opposite electric charge for our signal (\ellmp \psoftpm)
and use equal charge pairs (\ellpm \psoftpm) for background studies.  

We consider as signal all events where $\dsp\ell^-$ correlated production results 
in a peak near zero in \Mnu . Several processes contribute to the signal: 
(a) $\Bzb \rightarrow D^{*+} \ell^{-} \bar\nu_{\ell}$ decays (primary), 
(b) $\Bb \rightarrow D^{*+} n(\pi) \ell^- \bar{\nu}_{\ell}$ where the $D^{*+} n(\pi)$ may 
or may not originate from an excited charm state (\dstrstr), 
(c) $\Bzb \rightarrow D^{*+}\Db (\tau^-)$, $\Db(\tau^-) \rightarrow \ell^{-}X$ (cascade), and 
(d) $\Bzb \rightarrow D^{*+} h^-$ (fake), where the hadron ($h = \pi,K)$ is erroneously identified as a lepton
(in most of the cases, a muon). We also include radiative events, where one or more hard photons
are emitted by any charged particle, as described by the PHOTOS package~\cite{photos} in our simulation.
We define a signal region as $\Mnu > -2~$GeV$^{2}/c^4$. We use the sideband 
region, $-10 < \Mnu < -4~$GeV$^{2}/c^4$, for background studies.

\begin{figure}[!htb]
\begin{center}
\vspace*{-2.6cm}
\includegraphics[width=12cm]{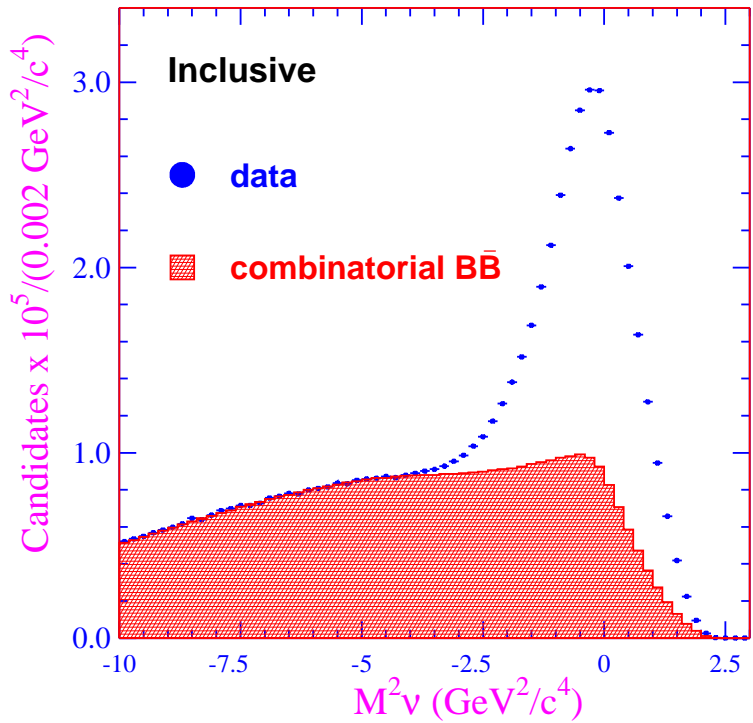}
\vspace*{-2cm}
\caption{The ${\cal M}_\nu^{\,2}$ distribution of the inclusive sample. The continuum background 
has been subtracted. The combinatorial background has been normalized to the data.} 
\label{fig:incl_yield}
\end{center}
\end{figure}

The background in the inclusive sample consists of continuum and combinatorial \BB\ events
(these last include also events where true \dsp\ and  $\ell^-$ from the two different $B$ mesons are combined).
We determine the number of signal events in our sample by fitting the \Mnu\ distribution in the 
interval $-10 <\Mnu< 2.5$ GeV$^2$/c$^4$. 
We perform the fit in ten bins of the lepton momentum
in order to reduce the sensitivity of the result to the details of the simulation.
We fix the continuum contribution to rescaled off-peak events, while we scale independently the number of 
signal events from primary, from \dstrstr\, and from combinatorial \BB\ predicted by the simulation.
We fix the contributions from cascade and fake decays, which account for about 3\% of the signal sample, 
to the Monte Carlo prediction. 
Figure~\ref{fig:incl_yield} shows the result of the fit in the \Mnu\ projection.
We then determine the number of signal events with $\Mnu > -2$ GeV$^2$/c$^4$,
as $N^{incl} = (2157.5 \pm 3.2 (stat) \pm 18.1 (syst)) \times 10^3$ events. The statistical
error includes the statistical uncertainties of the off-peak and of the simulated events.
The systematic error is discussed below.

We then look for $\Dz \to K^- \pip$ decays in the inclusive sample.
We consider all charged tracks in the event,  different from the \ellm and \psoft, matching 
the following criteria. We combine pairs of tracks with opposite electric charge, and we compute the
invariant mass $M_{K\pi}$, assigning the kaon mass to the track with charge opposite to the \psoft\ charge.
No identification requirement is applied to the $\pi^+$, while the $K^-$ satisfies its identification criterion.
We select events in the wide mass range $1.82 < M_{K\pi} < 1.91$ \gevcc, which contains more than 95\%
of our signal candidates. We then combine each \Dz\ candidate with the \psoft and look for signal events 
in the interval $142.4 < \Delta_{M} < 149.9$ \mevcc. We use the sideband region defined 
by  $153.5 < \Delta_{M} < 162.5$ \mevcc for background study.

\begin{figure}[!htb]
\begin{center}
\vspace*{-2.3cm}
\includegraphics[width=12cm]{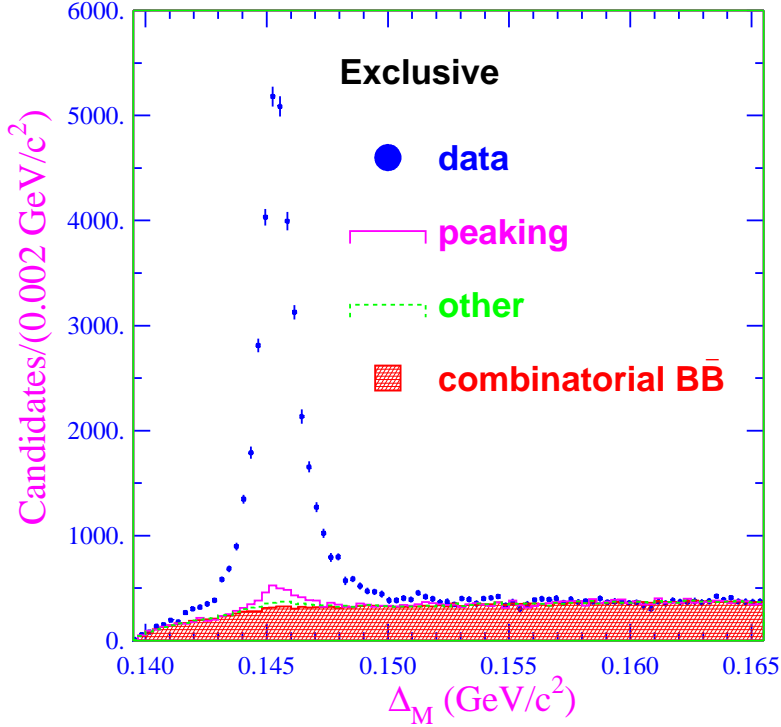}
\vspace*{-2cm}
\caption{The $\Delta_{M}$ mass distribution of the exclusive sample, after continuum subtraction.
The data are represented by points with its error. The hatched histogram shows the 
combinatorial \BB\ background. The other histograms show the uncorrelated peaking (solid histogram) 
and Cabibbo suppressed decays or other (dotted histogram), respectively.}
\label{fig:excl_yield}
\end{center}
\end{figure}

The exclusive sample consists of signal events, and of the following background sources:
continuum, combinatorial \BB, uncorrelated \dsp\ and Cabibbo-suppressed decays (uncorrelated peaking). 
We subtract the continuum using rescaled off-peak events selected with the same criteria
as the on-peak data as shown in Fig.~\ref{fig:excl_yield}. 
Combinatorial events are due to any combination of three tracks, 
in which at least one does not come from the \dsp. We determine their number from the
\BB\ Monte Carlo. We normalize the simulated events to the data in the \deltam\ sideband,
properly accounting for the small fraction of signal events (less than 1\%) contained in the sideband.
We verify that the background shape is properly described in the simulation using a sample of \dsp\ depleted
events, obtained as follows. First, we consider events in the \Mnu\ sideband. Then we 
consider ``wrong flavor'' events, where the candidate \Km charge is equal to the \psoft\ charge.
This sample contains more than 95\% combinatorial events in the signal region with a residual
peaking component from Cabibbo suppressed decays ($\Kp\Km$ and \pip \pim, see below).
After normalizing simulated events in the sideband, the number of events in the signal region 
is consistent with the data within the statistical precision of $\pm 1.3\%$.

Background from uncorrelated \dsp\ decays occurs when the \dsp\ and the \ellm\ originate from the 
two different $B$ mesons. These events exhibit a peak in \deltam\ but behave as combinatorial
background for \Mnu . We compute their number in the \Mnu\ sideband data, and rescale 
to the \Mnu\ signal region using the simulation events. 

Cabibbo suppressed decays  $\Dz \to \Km\Kp$ ($\Dz \to \pim \pip$) contribute to the peaking background, 
where one of the kaons (pions) is wrongly treated as a pion (kaon). Simulation shows that these events
peak in \deltam, while they exhibit a broad $M_{K\pi}$ distribution. Their amount is sizeably reduced by
a tighter requirement on $M_{K\pi}$. We subtract this background source using the simulation.
It should be noted that the contribution from Doubly Cabibbo Suppressed decays is negligible.

We finally obtain a sample of $N^{excl} = 31700 \pm 280$, where the error is statistical only. 
The detailed composition of the inclusive and exclusive data sets is presented in Table~\ref{tab:excl_results}.
\begin{table}[!htb]
\caption {Composition of the inclusive and exclusive samples.}
\begin{center}
\begin{tabular}{lrr} \hline \hline
Source              & Inclusive           &  Exclusive           \\ \hline 
Data                & $3887550 \pm 1970$  & $43920 \pm 210$      \\ 
Continuum           & $408960  \pm 1970$  & $2940  \pm 170$      \\
Combinatorial \BB\  & $1321250 \pm 580$   & $7410  \pm 50$       \\       
Peaking             & \dotfill            & $1370  \pm 80$       \\
Other               & \dotfill            & $510   \pm 10$       \\ \hline 
Yield               & $2157530 \pm 2850$  & $31700 \pm 280$      \\ \hline  \hline 
\end{tabular}
\end{center}
\label{tab:excl_results}
\end{table}

\section{Branching Fraction}

We compute the branching fraction as
\begin{eqnarray}
{\cal B}(\Dz \rightarrow \Km \pip) & = &  
             {N^{excl} \over N^{incl}} \times \frac{1}{\varepsilon_{(\Km \pip)}\beta} ,
\label{eq:branching_ratio}
\end{eqnarray}
where $\varepsilon_{(\Km \pip)} = (34.78 \pm 0.10)\%$ is the $\Dz$ reconstruction efficiency, and 
$\beta = 1.0496 \pm 0.0016$ is the analysis bias introduced by the partial reconstruction (see below).

The dominant contribution to the systematic uncertainty comes from the analysis bias. 
The bias factor $\beta$ is introduced to account for the fact that the efficiency of the inclusive 
event reconstruction is larger for two prong \Dz\ final states than for other events. 
Examining simulation, we find three independent bias sources. First,
the reconstruction of the soft-pion track is less efficient
for events with a high charged-tracks multiplicity, due to the larger density of hits near the \psoft\ track.
This accounts for a $(2.5 \pm 1.2)\%$ bias. A second contribution ($2.4 \pm 1.2\%$) is due to the cases in which
one of the charged tracks from the \Dz\ decay is preferred to the \psoft\ in forming the inclusive candidate,
and then the correct combination is lost. This is more frequent for large multiplicity \Dz\ decays,
because there are more charged tracks, and these have on average smaller momentum. Finally, the requirement 
on the minimal charged track multiplicity reduces by $\sim10\%$ the number of 0-prongs \Dz\ decays as
compared to two or more prongs. The contribution from this last effect is, however, less significant
($0.5 \pm 0.2\%)$. We take half of the deviation from unity as the systematic uncertainty on each source 
of the analysis bias. We consider all effects are independent and add these effects in quadrature to get 
a systematic uncertainty of $1.70\%$.

The main systematic uncertainty on $N^{incl}$ is due to the knowledge of the combinatorial \BB\ background 
(non-peaking combinatorial background). We perform a fit on the $\ellpm \psoftpm$ background control sample 
as we do on signal events. We take the RMS spread of the ratio between the data and the fit in the \Mnu\ projection 
resulting in a $0.75\%$ systematic uncertainty due to the combinatorial background (this choice is slightly 
more conservative than that adopted in~\cite{cleo98}). 
As first noticed in~\cite{cleo98}, the decays $\Bzb \to \ell^- \bar{\nu_\ell} D^+$, with 
$D^+ \to K^* \rho (\omega) \pi^+$, constitute a good-charged peaking background, because the charged pion 
is produced almost at rest in the \Dp\ decays. In order to estimate the systematic uncertainty due this
background, we vary its total fraction by $\pm 100\%$ in the \BB\ events in the Monte Carlo. 
Then we consider all the sources of systematic uncertainty which affect the shape of the signal
\Mnu\ shape. We vary by $\pm 30\%$ in turn the amount of events where at least one hard photon is radiated by
either the $\ell$ or the \psoft, or where the \psoft\ decays to a muon. We vary also by $\pm 30\%$
the fraction of cascade and fake decays, which are not determined by the fit. Finally, we vary in turn by $\pm 100\%$
the amount of events from each of the five sources constituting the \dstrstr\ samples (two narrow and two broad
resonant states, and non resonant \dsp\-pions). 
In each of the above studies, we repeat the fit, and take the variation in the result as 
the corresponding systematic uncertainty. 

To estimate the systematic uncertainty due to the background subtraction on $N^{excl}$, we first vary the number of
events from combinatorial background below the signal peak by the statistical uncertainty as described above. 
We then vary by $\pm 50\%$ the number of signal events contained in the sideband used for background
normalization, and we repeat the measurement. We vary the fraction of events from Cabibbo suppressed decays by the
uncertainty on the corresponding branching fraction, as reported in~\cite{pdg06}. As we determine the number of
uncorrelated peaking events from data, the corresponding systematic uncertainty is negligible.

Uncertainties of charged-track reconstruction, kaon identification, computation of $M(\Dz)$ and of \deltam 
affect the determination of the efficiency. The single charged-track reconstruction efficiency is determined with
$\pm 0.5\%$ precision. We then add linearly the error for the two tracks. The efficiency for $K^-$ 
identification is measured with $\pm 0.7\%$ systematic uncertainty from a large sample of 
$\dsp \to \Dz\pi^+, \Dz \to K^- \pi^+$ decays. When comparing a high purity signal sample, we observe 
a slight discrepancy between the shape of $M(\Dz)$ in the data and in the simulation. 
We compute a systematic error of $\pm 0.8\%$ due to the tuning of the simulated Monte Carlo events.

We compute the final relative systematic error of $2.43\%$ from the quadratic sum of 
all uncertainties listed above as shown in Table~\ref{tab:sys_errors}.
\begin{table}[!htb]
\caption{Summary of the relative systematic uncertainties of ${\cal B}(\Dz \rightarrow \Km \pip)$.}
\begin{center}
\begin{tabular}{llr} \hline \hline
Sample                 & Source                                   & $\delta({\cal B})/{\cal B}$ (\%)  \\ \hline
$N^{incl}$             & Analysis bias                            & $1.70$        \\ 
                       & Non-peaking combinatorial background     & $0.75$        \\
                       & Peaking combinatorial background         & $0.34$        \\
                       & Soft pion decays in flight               & $0.10$        \\ 
                       & Fake leptons                             & $0.08$        \\
                       & Cascade decays                           & $0.08$        \\
                       & Monte Carlo events shape                 & $0.08$        \\
                       & Continuum background                     & $0.05$        \\
                       & \dstrstr\ production                     & $0.02$        \\
                       & Photon radiation                         & $0.02$        \\  
$N^{excl}$             & Tracking efficiency                      & $1.0$         \\
                       & \Dz invariant mass                       & $0.8$         \\
                       & $K^-$ identification                     & $0.7$         \\ 
                       & Combinatorial background shape           & $0.3$         \\
                       & Combinatorial background normalization   & $0.27$        \\
                       & Other background                         & $0.1$         \\  \hline        
Total                  &                                          & $2.43$        \\  \hline \hline   
\end{tabular}
\end{center}
\label{tab:sys_errors}
\end{table}   
We cross check this result using four other alternative selections of the exclusive samples:
(1) we do not require that the $K^{-}$ to be identified, (2) we select \Dz\ events in 
a narrower ($\pm$25 MeV) band around the \Dz\ mass peak position, (3) we select \Dz\ events in 
a narrower band and require that the $K^-$ and the $\pi^+$ tracks originate from a common vertex, 
(4) combination of selections (2), (3) above, and require that the $K^{-}$ to be identified.
The background varies by a factor of 10, and the efficiency by about 30\%, from the looser to the 
tighter selection. All the results are consistent within the uncorrelated statistical and
systematic uncertainties.

\section{Summary}

In summary, we have measured the absolute branching fraction of $\Dz \rightarrow \Km \pip$ with partial
reconstruction of $\Bzb \rightarrow D^{*+} X \ell^{-} \bar\nu_{\ell}$. The preliminary result is
\begin{eqnarray}
{\cal B}(\Dz \rightarrow \Km \pip) & = & (4.025 \pm 0.038 \pm 0.098)\% ,
\label{eq:final_result}
\end{eqnarray}
which is consistent with the most precise results available to-date, and has a similar precision.

\section{Acknowledgments}

The author would like to thank all members of the \babar\ collaboration.
This work was supported in part by the U.S. Department of Energy 
contracts DE-AC02-76SF00515 and DE-FG05-91ER40622.

\end{document}